\def\year{2017}\relax
\documentclass[letterpaper]{article}
\usepackage{aaai17}
\usepackage{times}
\usepackage{helvet}
\usepackage{courier}
\usepackage{url}
\usepackage{float}

\usepackage{algorithm}
\usepackage{algorithmic}
\usepackage{comment}
\usepackage{caption}

\usepackage[utf8]{inputenc} 
\usepackage{url}            
\usepackage{booktabs}       
\usepackage{amsfonts}       

\usepackage{latexsym}
\usepackage{amsfonts}
\usepackage{amsmath}
\usepackage{amssymb}
\usepackage{color}
\usepackage{epsfig}
\usepackage{xspace}
\usepackage{graphicx}
\usepackage{subfigure}
\usepackage{epsfig}
\usepackage{multirow}
\usepackage{url}
\usepackage{balance}
\usepackage{wrapfig}
\usepackage{lipsum}

\newcommand{\ie}{\emph{i.e.,}\xspace}
\newcommand{\eg}{\emph{e.g.,}\xspace}
\newcommand{\wrt}{\emph{w.r.t.}\xspace}

\newcommand{\kw}[1]{{\ensuremath {\mathsf{#1}}}\xspace}

\newcounter{ccc}

\newcommand{\Jooin}{\mbox{\bf join}\ }
\newcommand{\Fetch}{\mbox{\bf fetch}\ }

\newcommand{\yago}{\kw{YAGO}}
\newcommand{\dbpedia}{\kw{DBpedia}}
\newcommand{\freebase}{\kw{Freebase}}

\newcommand{\ta}{\kw{TA}}

\newcommand{\sgta}{\kw{STAR}}
\newcommand{\mlsf}{\kw{L2P}-\kw{STAR}}

\newcommand{\pivot}{\kw{pivot}}

\newcommand{\ub}{\kw{UB}}
\newcommand{\lb}{\kw{LB}}

\newcommand{\quality}{F}
\newcommand{\NP}{\kw{NP}}

\begin{document}
%

\title{Learning to Speed Up Query Planning in Graph Databases}
\author{Mohammad Hossain Namaki$^\dagger$, F A Rezaur Rahman Chowdhury$^\dagger$ \\
{\bf \large Md Rakibul Islam, Janardhan Rao Doppa}, and {\bf Yinghui Wu} \\
\texttt{\{mnamaki, fchowdhu, mislam1, jana, yinghui\}@eecs.wsu.edu} \\
School of EECS, Washington State University, Pullman, WA, USA, 99163 \\
$\dagger$ Equal contribution from first two authors
}
\maketitle

\begin{abstract}
Querying graph structured data is a fundamental operation that enables important applications including knowledge graph search, social network analysis, and cyber-network security. However, the growing size of real-world data graphs poses severe challenges for graph databases to meet the response-time requirements of the applications. Planning the computational steps of query processing -- {\em Query Planning} -- is central to address these challenges. In this paper, we study the problem of learning to speedup query planning in graph databases towards the goal of improving the computational-efficiency of query processing via training queries. We present a {\em Learning to Plan} (L2P) framework that is applicable to a large class of query reasoners that follow the Threshold Algorithm (TA) approach. First, we define a generic search space over candidate query plans, and identify target search trajectories (query plans) corresponding to the training queries by performing an expensive search. Subsequently, we learn greedy search control knowledge to imitate the search behavior of the target query plans. We provide a concrete instantiation of our L2P framework for \sgta, a state-of-the-art graph query reasoner. Our experiments on benchmark knowledge graphs including \dbpedia, \yago, and \freebase show that using the query plans generated by the learned search control knowledge, we can significantly improve the speed of \sgta with negligible loss in accuracy.
\end{abstract}

\vspace{-1ex}
\section{Introduction}
\label{sec-intro}
\vspace{-.5ex}

Database technology has been successfully leveraged to improve the scalability and efficiency of artificial intelligence (AI) and machine learning (ML) algorithms \cite{db1,db2,db3,db4}. This paper focuses on the opposite direction of this successful cross-fertilization. We investigate ways to improve the computational-efficiency of querying databases by leveraging the advances from AI search, planning, and learning techniques. In this work, we study the following problem: {\em how can we automatically generate high-quality query plans, for minimizing the response time to find correct answers, by analyzing training queries drawn from a target distribution?}. This problem can be seen as an instance of  learning to speed up structured prediction \cite{SL,DP,JAIR14,MLS:AAAI14,CVPR15}.

\begin{figure}[h]
	\includegraphics[width=0.50\textwidth]{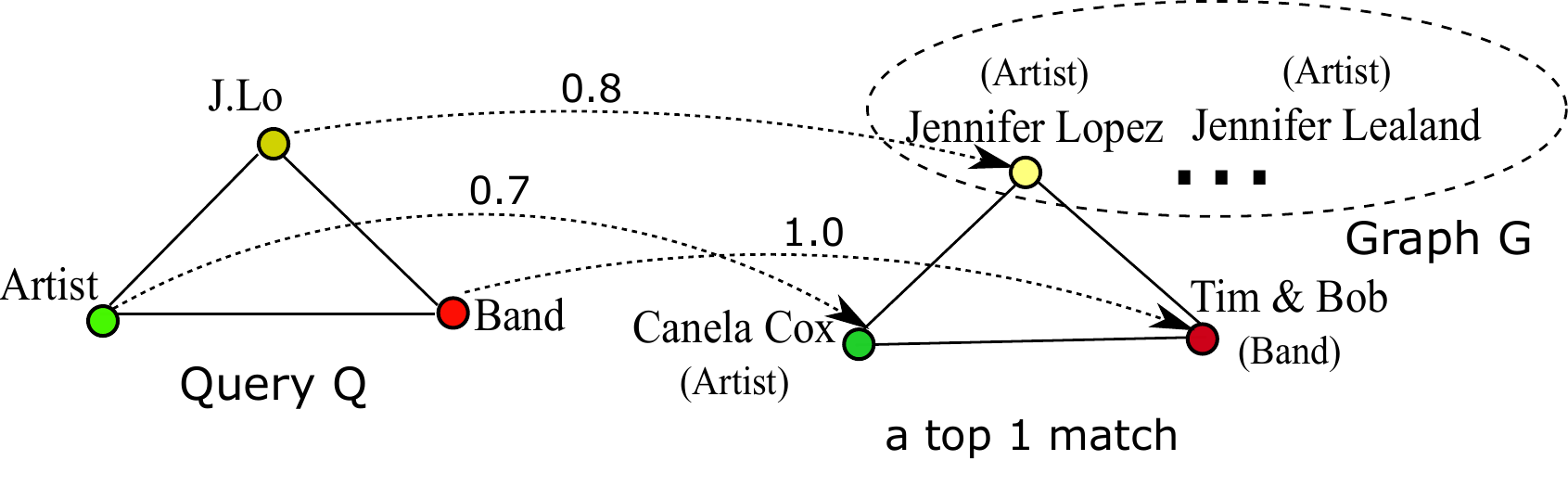}
\centering
\vspace{-4ex}
\caption{Illustration of Top-k graph querying problem.}
\label{fig:topkExample}
\end{figure}

Graph querying is the most primitive operation for information access, retrieval, and analytics over graph data that enables applications including knowledge graph search, and cyber-network security. We consider the problem of querying a graph database, where the input is a {\em data graph} and a {\em graph query}, and the goal is to find the answers to the given query by searching the data graph. For example, to detect potential threats, a network security expert may want to ``find communication patterns matching the attack pattern of a newly discovered Worm'' over a cyber network~\cite{chin2014predicting}. This natural language query is issued using a formal graph query language like SPARQL. Specifically, we study the general top-$k$ graph querying problem as illustrated in Example~\ref{fig:topkExample}.

\noindent {\bf Example 1.} Consider a graph query $Q$ on \dbpedia knowledge graph, shown in Figure \ref{fig:topkExample}. We want to find the artists who work with ``J.Lo" in a Band. Each of these ambiguous nodes in $Q$ can have excessive number of candidate matches. For example, there are 31,771 nodes with label ``Band" and 82,853 ``Artists" in \dbpedia. In addition, ``J.Lo" can match to different people whose first name starts with ``J" or lastname starts with ``Lo". While ``Jennifer Lopez" is the perfect match for this query, users may want to get other similar people like ``Jennifer Hudson", etc. However, the matching score for each candidate node can be different due to different node context. It is computationally expensive to first expand all of those matches to find the query structure and then rank them based on their matching scores.  

The growing size of real-world data graphs poses severe challenges for graph databases to meet the response-time requirements of the applications. Planning the computational steps of query processing -- {\em Query Planning} -- is central to address these challenges. In this work, we answer the main research question in the context of Threshold Algorithm (\ta) framework \cite{fagin2003optimal} for query processing that is widely popular in both graph and relational databases. The \ta framework works as follows. First, the given query is decomposed into sub-queries (e.g., star shaped queries for graph query). Subsequently, it follows a fixed {\em fetch-and-verify} strategy iteratively until a termination criterion (via estimates of lower and upper bound) is met. This framework has two major drawbacks: 1) It is very conservative in the upper bound estimation. This often leads to more computation without improvement in results; and 2) It applies static fetch and join policy for all the queries that will likely degrade its performance over a heterogeneous query set. In their seminal G{\"o}del prize-winning work, Fagin and colleagues suggested the usefulness of finding good heuristics to further improve the computational-efficiency of \ta framework as an interesting open problem \cite{fagin2003optimal}. Our work addresses this open problem with strong positive results. {\em To the best of our knowledge, this is the first work that tightly integrates learning and search to improve the computational-efficiency of query processing over graph databases}. 

We develop a general {\em Learning to Plan} (L2P) framework that is applicable to a large class of query reasoners that follow the \ta approach. First, we define a generic search space over candidate query plans, where the query plan of \ta framework can be seen as a greedy search trajectory. Second, we perform a computationally-expensive search (heuristic guided beam search) in this search space, to identify target search trajectories (query plans) corresponding to the training queries, that significantly improve over the computation-time of query plans from the \ta framework. Third, we learn greedy policies in the framework of {\em imitation learning} to mimic the search behavior of these target trajectories to quickly generate high-quality query plans. We also provide a instantiation of our L2P framework for \sgta, a state-of-the-art graph query reasoner and perform comprehensive empirical evaluation of \mlsf on three large real-world knowledge graphs. Our results show that \mlsf can significantly improve the computational-efficiency over \sgta with negligible loss in accuracy.

\section{Background}
\label{sec-pre}

In this section, we provide the background on graph query processing and the general \ta framework as applicable to both graph and relational databases.

\noindent {\bf Data Graph.} We consider a labeled and directed data graph $G$=$(V,E,\mathcal{L})$, with node set $V$ and edge set $E$. Each node $v\in V$ (edge $e\in E$) has a label $\mathcal{L}(v)$ ($\mathcal{L}(e)$) that specifies node (edge) information, and each edge represents a relationship between two nodes. In practice, $\mathcal{L}$ may specify heterogeneous attributes, entities, and relations~\cite{LuLWLW13}.

\noindent {\bf Graph Query.} We consider query $Q$ as a graph $(V_Q, E_Q, L_Q)$. Each {\em query node} in $Q$ provides information/constraints about an entity, and an edge between two nodes specifies the relationship posed on the two nodes. Formal graph query languages including SPARQL \cite{prud2006sparql}, Cypher, and Gremlin can be used to issue graph queries to a database such as Neo4j(\url{neo4j.com}). Since existing graph query languages are essentially subgraph queries, we can invoke a query transformer to work with different query languages \cite{kim2015taming}. Therefore, our work is general and is not tied to any specific query language.

\noindent {\bf Subgraph Matching.}  Given a graph query $Q$ and a data graph $G$, a match of $Q$ in $G$ is a mapping $\phi\subseteq V_Q\times V$, such that: 1) $\phi$ is an injective function, \ie for any pair of distinct nodes $u_i$ and $u_j$ in $V_Q$, $\phi(u_i)\neq\phi(u_j)$; and 2) for any edge $(v_i, v_j)\in E_Q$, $(\phi(v_i), \phi(v_j))\in E$. The match $\phi(Q)$ is a {\em complete} match if $|Q|=|\phi(Q)|$, where $|Q|$ denotes the size of $Q$, \ie the sum of the number of nodes and edges in $Q$ (similarly for $|\phi(Q)|$). Otherwise, $\phi(Q)$ is a {\em partial match}.

\noindent {\bf Matching Score.} Given $Q$ and a match $\phi(Q)$ in $G$, we assume the existence of a scoring function $F(\cdot)$ which computes, for each node $v\in V_Q$ (resp. each edge $e\in E_Q$), a matching score $F(\phi(v))$ (resp. $F(\phi(e))$). The matching score of $\phi$ is computed by a function $\quality(\phi(Q))$ as 
\vspace{-.5ex}
\begin{align}
\label{eq.score}
\quality(\phi(Q)) = \sum_{v \in V_Q}F(v, \phi(v))+\sum_{e\in E_Q}F(e,\phi(e))
\end{align}

One can use any similarity function such as acronym, abbreviation, edit distance etc. We adopt Levenshtein function~\cite{navarro2001guided} to compute node/edge label similarity, and employ the R-WAG's ranking function that incorporates both label and structural similarities \cite{6877688} to compute matching score $F(\cdot)$. 

\noindent {\bf Top-$k$ Graph Querying.} Given $Q$, $G$, and $\quality(\cdot)$, the top-$k$ subgraph querying problem is to find a set of $k$ matches $Q(G,k)$, such that for any match $\phi(Q) \notin Q(G,k)$, for all $\phi'(Q) \in Q(G,k)$, $\quality(\phi'(Q))$ $\geq \quality(\phi(Q))$.

\noindent {\bf Top-$k$ Search Paradigm.} Top-$k$ graph querying problem is known to be intractable (\NP-hard). The common practice in existing top-$k$ graph search techniques is to follow a conventional Threshold Algorithm (\ta) style approach~\cite{ding2014top,zeng2012top,ZouHWYHZ14,yangfast}.  

\begin{algorithm}[t]
\caption{\ta Framework for Graph Search}
\label{alg-sgta}
\footnotesize
\textbf{Input}: a graph query $Q$, a data graph $G$, integer $k$ \\
\textbf{Output}: top-$k$ match set $Q(G, k)$
\begin{algorithmic}[1]
\STATE Decompose $Q$ into a set of sub-queries $\mathcal{Q}$
\REPEAT
\STATE \Fetch $k$ partial matches for sub-queries in round-robin way
\STATE \Jooin to assemble new matches;
\STATE update lowerbound (\lb) and upperbound (\ub);
\UNTIL \ub $>$ \lb
\STATE \textbf{return} $Q(G, k)$
\end{algorithmic}
\end{algorithm}

\begin{figure}[t]
	\includegraphics[width=0.33\textwidth]{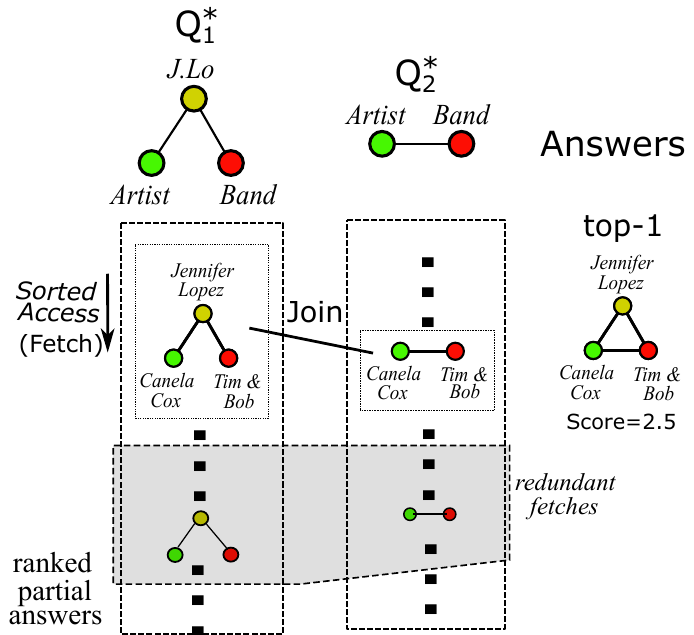}
\centering
\caption{Illustration of \ta via \sgta query processing.}
\label{fig:example2}
\end{figure}

\begin{figure}[t]
	\includegraphics[width=0.33\textwidth]{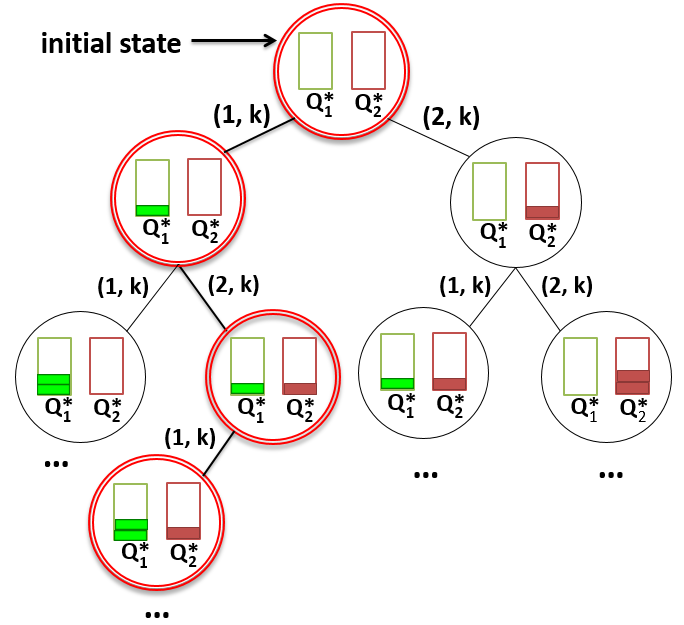}
\centering
\caption{Illustration of query plan from \ta framework.}
\label{fig:example4}
\end{figure}

\noindent {\bf \ta Framework for Query Processing.} The \ta framework was originally developed for relational databases \cite{fagin2003optimal}, and has been extended to graph databases as well. It consists of three main steps (see Algorithm \ref{alg-sgta}): {\bf Step 1:} {\em Query Decomposition.} A query $Q$ is decomposed into a set of sub-queries $\{Q_1, \ldots, Q_T\}$ without loss of generality. The procedure then initializes one list for each sub-query to store the partial answers. For relational databases, sub-queries can correspond to individual attributes \cite{fagin2003optimal}; and for graph databases, sub-queries can correspond to nodes or edges or stars \cite{yangfast}; {\bf Step 2:} {\em Partial Answer Generation.} For each sub-query $Q_i$, it performs an iterative exploration as follows. a) It computes and fetches $k$ partial matches of $Q_i$; and b) It verifies if the partial matches can be joined with other ``seen'' partial matches to be able to improve the top-$k$ matching score; and {\bf Step 3:} {\em Early Termination.} \ta dynamically maintains a) an lower bound ($LB$) as the smallest top-$k$ match score so far; and b) an upper bound ($UB$) to estimate the largest possible score of a complete match from unseen matches. If $UB < LB$, \ta terminates. Fig~\ref{fig:example2} illustrates \ta framework for the query $Q$ in Fig~\ref{fig:topkExample} and Fig~\ref{fig:example4} shows the query plan of \ta.

\noindent {\bf Limitations.} The main limitations of \ta framework are as follows: 1)  Applying fixed fetch-and-verify strategy for all the queries (\eg always fetch a fixed amount of partial answers) may not provide robust performance due to diversity in queries; 2) The performance of \ta framework highly depends on a good estimation of the upper bound. Enforcing a fixed estimation scheme can be loose, leading to 
``redundant fetches'' with little improvement in quality of answers (see Figure~\ref{fig:example2}). More adaptive stopping criterion is desirable; and 3) It is hard to adapt TA framework to different {\em resource constraints} (e.g., memory, response time, accuracy, I/O cost). 

Motivated by the above observations, we explore planning and learning techniques to {\em automatically induce adaptive data-driven search strategies} within the \ta framework to improve the efficiency of top-$k$ graph query processing. 

\section{Problem Setup}

We assume the availability of a training set of query-answer pairs $\left\{(Q, A^{\mathcal{P}})\right\}$ drawn from an unknown target distribution $\mathcal{D}$, where $Q$ is a graph query and $A^{\mathcal{P}}$ is the corresponding output answer produced by a \ta style algorithm $\mathcal{P}$ on data graph $G$. The accuracy of a query plan (sequence of computational steps for query processing) can be measured using a non-negative loss function $L$ such that $L(Q, \hat{A}, A^{\mathcal{P}})$ is the loss associated with processing a particular input query $Q$ to produce answer $\hat{A}$ when the correct answer is $A^{\mathcal{P}}$. The goal of learning is to quickly produce high-quality query plans for minimizing the response time to find correct answers for input queries drawn from the target distribution $\mathcal{D}$.

In this work, we formulate the graph query planning problem in a search-based learning framework. There are three key elements in this framework: 1) the {\em Search space} $\mathcal{S}_p$ whose states correspond to candidate computational states within the \ta framework; 2) the {\em Selection policy} $\Pi_{select}$ that is used to select the sub-query to fetch partial matches at each state; and 3) the {\em Fetching policy} $\Pi_{fetch}$ that is used to decide how many partial matches to fetch for the selected sub-query at each state.

\noindent {\bf Generic Search Space.} $\mathcal{S}_p$ is a 2-tuple $\left\langle I, A\right\rangle$, where $I$ is the initial state function, $A$ gives the set of possible actions in a given state. In our case, $s_0=I(Q)$, the initial state, corresponds to a state with one empty list for each sub-query decomposition of  $Q$ (say $T$ decompositions). $A(s_i)$ consists of actions that correspond to candidate computational steps (or search operators for query plans) at state $s_i$. Each action $a$ is of the form $(i, \delta)$ and corresponds to fetching $\delta$ additional partial matches for sub-query $i$, where $i \in \left\{1,2,\cdots,T\right\} \cup HALT$ and $\delta \in \left[\delta_{min}, \delta_{max}\right]$. When {\em HALT} action is chosen at state $s_i$ (terminal state), we stop the search and return the top-$k$ answers $\mathcal{A}(s_i)$.

We focus on greedy search (see Algorithm \ref{alg:greedy-inference}). The decision process for producing query plans corresponds to choosing a sequence of actions leading from the initial state using both $\Pi_{select}$ and $\Pi_{fetch}$,  until $\Pi_{select}$ selects the {\em HALT} action (terminal state). $\Pi_{select}$ and $\Pi_{fetch}$ are parametrized by feature functions $\psi_1$ and $\psi_2$. We want to learn the parameters of both $\Pi_{select}$ and $\Pi_{fetch}$ using the training queries, with the goal of quickly producing query plans for minimizing the response time to produce correct answers on unseen queries drawn from $\mathcal{D}$.

\section{Learning to Plan (L2P) Framework}

Inspired by the recent success of imitation learning approaches for solving sequential decision-making tasks \cite{nlp2,mip1,nlp3,SP}, we formulate and solve the problem of learning policies to produce high-quality query plans in the framework of imitation learning.

\begin{algorithm}[bt]
\footnotesize
\caption{L2P for \ta Framework}
\textbf{Input}: $Q = $ graph query,
$G = $ data graph,
$\Pi_{select} = $ selection policy,
$\Pi_{fetch} = $ fetching policy
\label{alg:greedy-inference}
\begin{algorithmic}[1]
\STATE $s \leftarrow I(Q)$ {\em // initial state}
\STATE $\text{TERMINATE}\leftarrow False$
\WHILE{\textbf{not} \mbox{TERMINATE}}
\STATE $a_{select} \leftarrow \Pi_{select}(s)$ {\em // select the sub-query} 
\IF{$a_{select} == HALT$}
\STATE $\text{TERMINATE} = True$
\ELSE
\STATE $a_{fetch} \leftarrow \Pi_{fetch}(s, a_{select})$ {\em // how many to fetch}
\STATE $s \leftarrow$ Apply $(a_{select}, a_{fetch})$ on $s$
\ENDIF
\ENDWHILE
\STATE \textbf{return} Top-$k$ answers $\mathcal{A}(s)$ corresponding to $s$
\end{algorithmic}
\end{algorithm}

\noindent {\bf Overview of Imitation Learning.} In traditional imitation learning, expert demonstrations are provided as training data (e.g., demonstrations of a human expert driving a car), and the goal of the learner is to learn to imitate the behavior of an expert performing a task in a way that generalizes to similar tasks or situations. Typically this is done by collecting a set of trajectories of the expert's behavior on a set of training  tasks. Then supervised learning is used to find a policy that can replicate the decisions made on those trajectories. Often the supervised learning problem corresponds to learning a classifier or regressor to map states to actions, and off-the-shelf tools can be used.

The two main challenges in applying imitation learning framework to query planning are: 1) Obtaining a high-quality oracle policy that will provide the supervision for the imitation learning process ({\em oracle construction}); and 2) Learning search control policies that can make search decisions in a computationally-efficient manner ({\em fast and accurate decision-making}). We provide solutions for these two challenges below.
\vspace{-1ex}
\subsection{Oracle Construction}
\vspace{-.5ex}

In this section, we describe a generic procedure to compute high-quality query plans within the \ta framework. We will start by defining the needed terminology.

\noindent {\bf Definition 1.} For a given query-answer pair $(Q, A^{\mathcal{P}})$, a state $s$ in the search space $\mathcal{S}_p$ is called a {\em terminal state} if the corresponding top-$k$ answers $\mathcal{A}(s)$ is same as $A^{\mathcal{P}}$ (output of \ta style algorithm $\mathcal{P}$ for query $Q$). 

\noindent {\bf Definition 2.} For a given query $Q$, a sequence of actions from initial state to terminal state, $(s_{0}, a_{0}),\cdots,(s_{N}, \varnothing)$, is called a {\em target query plan} (TQP), where $s_{0}$ is the initial state and $s_{N}$ is the terminal state.

\noindent {\bf Definition 3.} For a given query $Q$, the {\em quality of a TQP} is defined as the computation time taken to execute the query plan to produce the top-$k$ answers. In other words, quality corresponds to speedup \wrt computation time of \ta style algorithm $\mathcal{P}$ on the same query.

The goal of oracle construction is to compute high-quality target query plans \cite{JMLR14} for each training query-answer pair $(Q, A^{\mathcal{P}})$. A naive approach would be to perform depth-bounded exhaustive search in search space $\mathcal{S}_p$ instantiated for each training query. Since the naive approach is impractical, we resort to heuristic guided beam search and define effective heuristics to make the search efficient.

\noindent {\bf Heuristics.} We define multiple heuristics that can potentially guide the search to uncover high-quality target query plans. These heuristics are defined as a function of the given search state $s$ and the terminal state $s^{\mathcal{P}}$ of the \ta style algorithm $\mathcal{P}$. We define the following three concrete heuristics noting that our approach allows to add additional heuristics as needed: {\bf H1)} Total computation time to reach state $s$ from initial state $s_0=I(Q)$ normalized \wrt time taken by the \ta style algorithm $\mathcal{P}$; {\bf H2)} The cost (or score) of answer $\mathcal{A}(s)$ normalized \wrt the score of $A^{\mathcal{P}}$ \cite{Arai:2007:AMT:1325851.1325954}; and {\bf H3)} The cumulative difference between the sizes of lists (with partial matches) for each sub-query from states $s$ and $s^{\mathcal{P}}$ normalized \wrt the total size of all lists for $s^{\mathcal{P}}$. Intuitively, H1, H2, and H3 correspond to speed, accuracy, and distance to goal in terms of most primitive action respectively. We seek to balance the accuracy and computational cost by appropriately weighting the heuristics.

\noindent {\bf Target Query Plan Computation via Beam Search.} We propose to combine multiple heuristics by their weighted linear combination: $H(s)$=$\sum_{i=1}^{m} w_i \cdot H_i(s)$. For a given weight vector $w \in \Re^{m}$, the heuristic function is fully specified. For a given training query-answer pair $(Q, A^{\mathcal{P}})$, we perform breadth-first beam search with beam width $b$ starting at initial state $s_0=I(Q)$, until we uncover a terminal state (i.e., $\mathcal{A}(s)$ is same as $A^{\mathcal{P}}$). The sequence of actions leading from $s_0$ to terminal state $s$ is identified as the target query plan for $Q$ (see Algorithm \ref{alg:Oracle-Procedure}). 

\noindent {\bf Computing Weights via Bayesian Optimization.} We don't know the appropriate weights $w_1, w_2,\cdots,w_m$ that will improve the effectiveness of the weighted heuristic function $H$. We define the value of a candidate weight vector $w \in \Re^m$ over a set of training queries $\mathcal{T}$, $\mathcal{V}(w, \mathcal{T})$, as the average quality (speedup) of the identified target query plans (via Algorithm \ref{alg:Oracle-Procedure}) for all queries in $\mathcal{T}$ with the corresponding heuristic $H$. Our goal is to find the weight vector $w \in \Re^m$ that maximizes $\mathcal{V}(w, \mathcal{T})$: 
\begin{equation}
w^*=\mbox{arg}\,max_{w \in \Re^m} \, \mathcal{V}(w, \mathcal{T})
\end{equation}

The main challenge in solving this optimization problem is that evaluating the value of a candidate weight vector is computationally-expensive. In recent years, Bayesian Optimization (BO) \cite{BO} has emerged as a very popular framework for solving optimization problems where the function evaluations are computationally expensive (e.g., hyper-parameter tuning of machine learning algorithms). Therefore, we propose to find the best weights via BO tools. In short, BO algorithms build a statistical model based on the past function evaluations; and execute a sequential decision-making process that intelligently explores the solution space guided by the statistical model, to quickly reach a near-optimal solution. 

\begin{algorithm}[bt]
\footnotesize
\caption{Target Query Plan Computation}
\textbf{Input}: $(Q, A^{\mathcal{P}}) = $ query and answer pair, $(I,A) = $ Search space, $b = $ beam width, $H = $ heuristic function 
\label{alg:Oracle-Procedure}
\begin{algorithmic}[1]
\STATE $B \leftarrow s_0=I(Q)$ {\em // Initial state}
\STATE $\text{TERMINATE}\leftarrow False$
\WHILE{\textbf{not} \mbox{TERMINATE}}
\STATE $C \leftarrow \emptyset$ {\em // Candidate set}
\FOR{each state $s \in B$}
\IF{$\mathcal{A}(s) == A^{\mathcal{P}}$}
\STATE $\text{TERMINATE} = True$
\STATE $s^* = s$
\ELSE
\STATE Expand $s$ and add all next states to $C$ {\em // Expansion}
\ENDIF
\ENDFOR
\STATE $B \leftarrow$ Top-$b$ scoring states in $C$ via heuristic {\em // Pruning}
\ENDWHILE
\STATE \textbf{return} state-action pair sequence from $s_0$ to $s^*$
\end{algorithmic}
\end{algorithm}

\subsection{Learning Greedy Policies via Imitation Learning}

Our goal is to learn a greedy policy $(\Pi_{select}, \Pi_{fetch})$ that maps states to actions in order to imitate the target query plans computed via oracle construction. We assume that for any training query-answer pair $(Q,A^{\mathcal{P}})$, we can get the oracle query plan $(s^{*}_{0}, a^{*}_{0}),(s^{*}_{1}, a^{*}_{1}),\cdots,(s^{*}_{N}, \varnothing)$ via Algorithm \ref{alg:Oracle-Procedure}, where $s^{*}_{0}$ is the initial state and $s^{*}_{N}$ is the terminal state. The goal is to learn the parameters of $\Pi_{select}$ and $\Pi_{fetch}$ such that at each state $s^{*}_{i}$, $a^{*}_{i} \in A(s^{*}_{i})$ is selected.

\vspace{-1ex}
\begin{algorithm}[H]
\caption{Learning Greedy Policy via Exact Imitation}
\footnotesize
\textbf{Input}: $\mathcal{T} = $ Training data
\label{alg:exact-imitation-learning}
\begin{algorithmic}[1]
\STATE Initialize the set of classification examples $\mathcal{D}_1 =\emptyset$
\STATE Initialize the set of regression examples $\mathcal{D}_2 =\emptyset$
\FOR{each training example $(Q, A^{\mathcal{P}}) \in \mathcal{T}$}
\STATE Compute the target query plan 
$(s^{*}_{0}, a^{*}_{0}),\cdots,(s^{*}_{N}, \varnothing)$ 
\FOR{each search step $t=0$ to $N$}
\STATE Generate classification example $C_t$ and regression example $R_t$ to imitate $(s^{*}_{t}, a^{*}_{t})$
\STATE Aggregate training data: $\mathcal{D}_1=\mathcal{D}_1 \cup C_t$ and $\mathcal{D}_2=\mathcal{D}_2 \cup R_t$
\ENDFOR
\ENDFOR
\STATE $\Pi_{select} = $ Classifier-Learner($\mathcal{D}_1$) 
\STATE $\Pi_{fetch} = $ Regression-Learner($\mathcal{D}_2$) 
\STATE \textbf{return} greedy policy $(\Pi_{select}, \Pi_{fetch})$
\end{algorithmic}
\end{algorithm}

\noindent {\bf Exact Imitation Approach.} At each state $s^{*}_{t}$ on the target path of a training example $(Q,A^{\mathcal{P}})$, we create one classification example with $\psi_1(s^{*}_{t})$ as input and $a^{*}_{t}(1)$ (selection action) as output; and one regression example with $\psi_2(s^{*}_{t}, a^{*}_{t}(1))$ as input and $a^{*}_{t}(2)$ (fetching action) as output. The sets of aggregate classification and regression imitation examples collected over all the training queries are then fed to a classifier and regression learner pair, to learn the parameters of $\Pi_{select}$ and $\Pi_{fetch}$ (see Algorithm \ref{alg:exact-imitation-learning}). This reduction allows us to leverage powerful off-the-shelf classification and regression learners. In theory and practice, policies learned via exact imitation can be prone to error propagation: errors in the previous state contributes to more errors. 

\noindent {\bf DAgger Algorithm.} DAgger is an advanced imitation approach \cite{DAGGER} that generates additional training data so that the learner is able to learn from its mistakes. It is an iterative algorithm, where each iteration adds imitation data to an aggregated data set. The first iteration follows the exact imitation approach. After each iteration, we learn policy $(\Pi_{select}, \Pi_{fetch})$ using the current data. Subsequent iterations perform query planning using the learned policy to generate a trajectory of states for each training query. At each decision along this trajectory, we add a new imitation example if the search decision of the learned policy is different from the oracle policy. In the end, we select the best policy over all the iterations via performance on validation data. To select among several imperfect policies with varying speed and accuracy performance, we pick the policy with the highest accuracy. This principle is aligned with our learning objective.

\subsection{Handling General Query Planning}

In this section, we provide some discussion on ways to extend our L2P framework to more general query planning going beyond the \ta style query processing. 

There are three key elements in the L2P framework: 1) Search space over query plans; 2) Policy for producing query plans; and 3) Oracle query plans to drive the learning process. The search operators for query plans are specific to each query processing approach. We need to consider additional search operators (e.g., different $\delta$ values in the \ta framework) to be able to construct high-quality candidate query plans in the search space. The form of the policy will depend on the search operators or actions at search states (e.g., classifier-regressor pair to select the sub-query and number of partial matches to fetch in the TA framework). For computing the oracle plans needed to learn the policy, heuristic-guided beam search is a generic approach, but we need to define effective heuristics as applicable for the given query evaluation approach. A more generic off-the-shelf alternative is to consider {\em Approximate Policy Iteration} (API) algorithm \cite{API}. API starts with a default policy and iterates over the following two steps. {\bf Step 1:} Generate trajectories of current rollout policy from initial state; and {\bf Step 2:} Learn a fast approximation of rollout policy via supervised learning to induce a new policy. Indeed, each iteration of API can be seen as imitation learning, where trajectories of current rollout policy correspond to expert demonstrations and the new policy is induced using the exact imitation algorithm \cite{PC}. 

\section{L2P Instantiation for \sgta}

In this section, we provide a concrete instantiation of our L2P framework for \sgta~\cite{yangfast}, a state-of-the-art graph query reasoner based on the \ta framework. 

\noindent {\bf Overview of \sgta Query Processing.} \sgta is an instantiation of \ta framework, where the graph query is decomposed into a set of star-shaped queries. It was shown both theoretically and empirically that star decomposition provides a good trade-off between sub-query evaluation cost and the number of candidate partial answers. Figure~\ref{fig:example2} illustrates \sgta query processing to find the top-$1$ answer for the query $Q$ shown in Figure~\ref{fig:topkExample}. It first decomposes $Q$ into two stars ($Q^{*}_1$ and $Q^{*}_2$), i.e., graphs with a unique \pivot and one or more adjacent nodes. It then fetches partial answers for each star query in a sorted manner, and joins the partial answers whenever possible, until it finds a complete match and termination criteria is met.

\noindent {\bf Search Space.} Suppose the given graph query $Q$ is decomposed into a set of star-queries $\{Q^*_1, \ldots, Q^*_T\}$ without loss of generality. 
Recall that each action $a$ (i.e., search operator for query plan) is of the form $(i, \delta)$ and corresponds to fetching $\delta$ additional partial matches for star query $i$, where $\delta \in \left[\delta_{min}, \delta_{max}\right]$. We employed $\delta_{min}=10$ and $\delta_{max}=200$ (candidate number of partial matches to fetch), and considered candidate choices in the multiples of $\delta_{min}$, i.e., $\frac{\delta_{max}}{\delta_{min}}$ discrete values. This choice is mainly driven by the computational complexity of finding oracle query plans via breadth-first beam search. The branching factor at each search step is $b \times T \times \frac{\delta_{max}}{\delta_{min}}$, where $b$ is the beam width. 

\noindent {\bf Features.} To drive the learning process, we define feature functions $\psi_1$ and $\psi_2$ over search states. Our features can be categorized into three groups: 1) {\em Static features} that are computed from the query topology, decomposed star queries, and initial partial matches. Some examples include the number of nodes and edges in each star query, the number of candidate nodes in data graph, and the number of joinable nodes in a star decomposition; 2) {\em Ranking features} that are computed from the lower bound of current top-$k$ answers and upper bound for each star query; and 3) {\em Context features} that are computed based on the current state of \mlsf like the selected star query and the total number of fetches for each star query. 
See https://goo.gl/MD51S2 for complete details of features. Alternatively, we can employ a deep learning model to learn appropriate representation.

\section{Experiments and Results}
\label{sec-experiment}

In this section, we present our empirical evaluation.

\subsection{Experimental Setup}  

\noindent {\bf Datasets.} We employ three real-world open knowledge graphs: 1) \yago (\url{mpi-inf.mpg.de/yago}) contains $2.6M$ entities (\eg people, companies, cities), $5.6M$ relationships, and $10.7M$ nodes and edge labels, extracted from several knowledge bases including Wikipedia;
2) \dbpedia (\url{dbpedia.org}) consists of $3.9M$ entities, $16.8M$ edges, and $14.9M$ labels; and 3) \freebase (\url{freebase.com}), 
a more diversified knowledge graph that contains $40.3M$ entities,  $180M$ edges, and $81.6M$ labels. 

\begin{table}[]
\centering
\small
\begin{tabular}{|c|c|c|}
\hline
Yago & DBPedia & Freebase \\ \hline
590s   & 680s   & 1120s      \\ \hline
\end{tabular}
\caption{Average runtime of Algorithm~\ref{alg:Oracle-Procedure} with $b$=10.}
\label{tbl:oracleSearchTime}
\end{table}

\begin{table*}[t]
\centering
\small
\begin{tabular}{l|c|c|c|c|c|c|c|c|c|}
\cline{2-10}
\multirow{2}{*}{}                            & \multicolumn{3}{c|}{\yago}                                                                         & \multicolumn{3}{c|}{\dbpedia}                                                                     & \multicolumn{3}{c|}{\freebase}                                                                     \\ \cline{2-10} 
                                             & \multicolumn{1}{l|}{speedup} & \multicolumn{1}{l|}{accuracy} & \multicolumn{1}{l|}{run-time (ms)} & \multicolumn{1}{l|}{speedup} & \multicolumn{1}{l|}{accuracy} & \multicolumn{1}{l|}{run-time (ms)} & \multicolumn{1}{l|}{speedup} & \multicolumn{1}{l|}{accuracy} & \multicolumn{1}{l|}{run-time (ms)} \\ \hline
\multicolumn{1}{|l|}{\sgta}                     & 1.00                          & 100\%                         & 1925.78                            & 1.00                          & 100\%                         & 6401.48                           & 1.00                          & 100\%                         & 13932.26                           \\ \hline
\multicolumn{1}{|l|}{Oracle}                 & 4.53                          & 100\%                         & 757.70                             & 5.62                          & 100\%                         & 1192.26                           & 62.53                         & 100\%                         & 1478.60                            \\ \hline
\multicolumn{1}{|l|}{Random}         & 3.37                         & 61\%                          & 1033.6                             & 10.40                        & 66\%                          & 549.30                            & 40.20                        & 54\%                          & 2555.59                            \\ \hline
\multicolumn{1}{|l|}{\mlsf (Exact)} & 3.66                         & 93\%                          & 764.00                            & 5.00                          & 97\%                          & 1310.26                          & 47.69                         & 95\%                          & 1550.03                            \\ \hline
\multicolumn{1}{|l|}{\mlsf (DAgger)} & 3.71                         & 94\%                          & 804.61                             & 5.00                         & 97\%                          & 1310.26                            & 47.71                        & 96\%                          & 1508.03                            \\ \hline
\end{tabular}
\caption{Speedup, accuracy, and query run-time results comparing \sgta, Oracle, Random, and \mlsf.}
\label{tbl:highlevel}
\end{table*}

\begin{table*}[t]
\small
\centering
\begin{tabular}{l|c|c|c|c|c|c|c|c|c|c|}
\cline{2-11}
\multirow{3}{*}{}   &\multicolumn{5}{c|}{\dbpedia}   & \multicolumn{5}{c|}{\freebase}  \\ 
\cline{2-11} 
 & \multicolumn{2}{c|}{Analytical Ops} & \multicolumn{3}{c|}{Computation Time (ms)}  & 
\multicolumn{2}{c|}{Analytical Ops} & \multicolumn{3}{c|}{Computation Time (ms)}   
\\ \cline{2-11} 
  & \# Fetch    & \# Join   & Fetch  & Join & ML-Overhead  
	& \# Fetch   & \# Join   & Fetch & Join & ML-Overhead 
	\\ \hline
\multicolumn{1}{|l|}{\sgta}   & 3,876             & 13,106          & 3,101.51   & 3,283.58  & N/A              & 5,732             & 20,048          & 3,868.65   & 10,044.39 & N/A              \\ \hline
\multicolumn{1}{|l|}{Oracle}   & 28                & 43              & 1,208.89   & 6.32      & N/A              & 35                & 65              & 1,487.12   & 7.64      & N/A              \\ \hline
\multicolumn{1}{|l|}{Random} & 7                 & 13              & 549.29      & 0.05        & N/A            & 8                 & 15              & 2,554.76    & 0.35         & N/A           \\ \hline
\multicolumn{1}{|l|}{\mlsf}  & 80                & 134             & 1,238.89    & 49.69       & 23.01          & 133               & 260             & 1,440.39    & 46.52        & 18.34         \\ \hline
\end{tabular}
\caption{Statistics of different analytical operations (e.g., fetch and join calls) and the corresponding computation time.}
\label{tbl:lowLevel}
\end{table*}

\noindent{\bf Query Workload.} We developed a query generator to produce both training and testing queries following the {\em DBPSB} benchmark \cite{MOR+11}. We first generate a set of query templates, each has a topology sampled from a graph category, and is assigned with labels (specified as ``entity types'') sampled from top $20\%$ most frequent labels in the real-world graphs. We created $20$ templates to cover common entity types, and generated a total of $2K$ queries by instantiating the templates. We employ 50\% queries for training, 20\% for validation, and 30\% for testing respectively.

\begin{table*}[t]
\small
\centering
\begin{tabular}{lc|c|l|l|l|l|l|}
\cline{3-8}
\multicolumn{2}{l}{\multirow{3}{*}{}}                      & \multicolumn{6}{|c|}{Expansion}                                                                                                                                                         \\ \cline{3-8} 
\multicolumn{2}{l}{}                                       & \multicolumn{2}{|c|}{\yago}                                        & \multicolumn{2}{c|}{\dbpedia}                             & \multicolumn{2}{c|}{\freebase}                            \\ \cline{3-8} 
\multicolumn{2}{l}{}                                       & \multicolumn{1}{|c|} \sgta                              & \multicolumn{1}{c|}{\mlsf} & \multicolumn{1}{c|}{\sgta} & \multicolumn{1}{c|}{\mlsf} & \multicolumn{1}{c|}{\sgta} & \multicolumn{1}{c|}{\mlsf} \\ \hline
\multicolumn{1}{|l|}{\multirow{2}{*}{Selection}} & \sgta    & \multicolumn{1}{l|}{(1, 100\%)}   & (1.86, 100\%)                & (1, 100\%)                & (2.19, 100\%)                & (1, 100\%)                & (3.36,100\%)                \\ \cline{2-8} 
\multicolumn{1}{|l|}{}                           & \mlsf & \multicolumn{1}{l|}{(4.08, 87\%)} & (3.71, 94\%)                 & (5.81, 90\%)              & (5.00, 97\%)                 & (34.04, 91\%)             & (47.71, 96\%)                \\ \hline
\end{tabular}
\caption{Results of ablation analysis.}
\label{tbl:ablation}
\end{table*}

\noindent {\bf \sgta Implementation.} We implemented the \sgta query processing framework \cite{yangfast} in Java using the Neo4j (\url{neo4j.com}) graph database system.

\noindent {\bf Oracle Policy Implementation.} We performed heuristic guided breadth-first beam search with different beam widths $b=\left\{1,5,10,20,50\right\}$ to compute high-quality target query plans for each training query (see Algorithm \ref{alg:Oracle-Procedure}). BayesOpt software \cite{JMLR:v15:martinezcantin14a} was employed to find the weights of heuristics with expected improvement as the acquisition function. We did not see noticeable performance improvement beyond 100 iterations. Since we didn't get significant speedup improvements beyond $b=10$, we employed target query plans (aka Oracle policy) obtained with $b=10$ for all our training and testing experiments.

\noindent {\bf \mlsf Implementation.} We employed XGBoost~\cite{XGBOOST}, an efficient and scalable implementation of functional gradient tree boosting for classification and regression, as our base learner. All hyper-parameters (boosting iterations, tree depth, and learning rate) were automatically tuned based on the validation data using BayesOpt \cite{JMLR:v15:martinezcantin14a}, a state-of-the-art BO software. \mlsf (Exact) and \mlsf (DAgger) corresponds to policy learning via exact imitation and DAgger algorithms respectively. We performed 5 iterations of DAgger with $\beta=0.8$ and exponential decay. We selected the policy with the highest accuracy on the validation data. \mlsf needs to additionally  store the learned policy (classifier and regressor pair). However, this overhead is negligible when compared to the memory usage of Neo4j.

\noindent {\bf Code and Data.} All the code and data is available on a GitHub repository: https://goo.gl/MD51S2.

\noindent {\bf Evaluation Metrics.} We evaluate \sgta, Oracle, and \mlsf using two metrics: 1) {\em Speedup.} For a given query $Q$, the response time of an algorithm $\mathcal{P}$, denoted as {\tt Time}$(\mathcal{P}, Q)$, refers to the total CPU time taken from receiving the query to finding the top-$k$ answer. The speedup factor $\tau(\mathcal{P},Q)$ is computed as the ratio of {\tt Time}$(\sgta, Q)$ to {\tt Time}$(\mathcal{P}, Q)$; and 2) {\em Accuracy.} We employ Levenshtein distance function~\cite{navarro2001guided} to compute node/edge label similarity, and  R-WAG's ranking function\cite{6877688} to compute the matching score $F(\cdot)$. For a given query $Q$, the accuracy of an algorithm $\mathcal{P}$ is defined as the ratio of the matching score of correct answer $A^{\mathcal{P}}$ to the matching score of predicted answer $\hat{A}$. 

We conducted all our experiments on a Windows server with 3.5 GHz Ci7 CPU and 32GB RAM configuration. All experiments are conducted 3 times and averaged results are presented. We report the average metrics (speedup, accuracy, and computation time) over all testing queries.

\subsection{Results and Analysis} 

\noindent {\bf Oracle Policy Computation Time.} Table \ref{tbl:oracleSearchTime} shows the average time to compute the oracle query plan via beam search with beam width $b$=10 (see Algorithm~\ref{alg:Oracle-Procedure}). The runtime is high and it increases for denser graphs (e.g., \freebase). Hence, we cannot use this computationally expensive procedure in real-time and need L2P to quickly generate high-quality query plans.

\noindent {\bf \sgta vs. Oracle.} To find out the overall room for improvement via L2P framework, we compare \sgta with the oracle policy. Recall that the oracle policy for a given query $Q$, corresponds to the target query plan obtained by heuristic-guided breadth-first beam search, and relies on the knowledge of correct answer $A^{\mathcal{P}}$. Table \ref{tbl:highlevel} shows the speed, accuracy, and run-time; and Table~\ref{tbl:lowLevel} shows the statistics of different analytical operations (e.g., fetch and join calls) and the corresponding computation time. 

We make the following observations: 1) There is significant room for improving \sgta via L2P framework as noted by speedup of oracle policy for different datasets (4.53 for \yago, 5.62 for \dbpedia, and 62.53 for \freebase); 2) The speedup of oracle policy is higher for large and dense data graphs, \eg \freebase. Indeed, for dense graphs, we expect more candidate matches for each star-query, which can make the \sgta very slow and \mlsf would be more beneficial. In fact, there is a growing evidence that the real-world graphs become denser over time and follow a power-law pattern \cite{leskovec2007graph}; and 3) The number of fetch/join calls can be reduced by two orders of magnitude by following the plan from oracle policy. The number of fetch calls show the number of search steps to reach the {\em correct} terminal state, which is much smaller for oracle when compared to \sgta.  As pointed out earlier, one of the major drawbacks of \ta style \sgta is that it fetches many ``useless'' partial answers that do not contribute to top-$k$ answers. The performance of oracle policy shows that it is possible to significantly improve \sgta if the learner can successfully imitate the oracle plans. 

\noindent {\bf \mlsf vs. Oracle.} We compare \mlsf with the oracle policy to understand how well the learner is able to mimic the search behavior of oracle. We make the following observations from Table~\ref{tbl:highlevel}. The speed and accuracy of \mlsf in general is very close to the oracle policy across all the datasets. \mlsf loses at most $6\%$ accuracy and significantly improves the speed when compared to \sgta. For example, \mlsf improves the speed of \sgta by $47$ times with an accuracy loss of $5\%$ for \freebase. \mlsf has to learn when to stop the search (i.e., select {\em HALT} action). If \mlsf stops the search early, it will lose accuracy. Similarly, it will lose speed when stopping of the search is delayed. From Table~\ref{tbl:lowLevel}, we can see that the overhead of \mlsf to make search decisions (i.e., computing features and executing the classifier/regressor) is very small when compared to the query processing time ($2\%$ of query run-time on an average). 

We saw small performance improvement in \mlsf by training with DAgger when compared to training with exact-imitation (except for \dbpedia). This is due to our principle of selecting among multiple imperfect policies: pick the policy with highest accuracy. We were able to uncover policies with higher speedup over exact imitation, but their accuracy was relatively low.

\noindent {\bf Random Policy.} We also compared with a baseline, where both $\Pi_{select}$ and $\Pi_{fetch}$ are random. Table~\ref{tbl:highlevel},~\ref{tbl:lowLevel} show the results averaged over 20 different runs. Random policy has better speedup than \sgta, but with a significant drop in accuracy as it selects the HALT action prematurely.

\noindent {\bf Ablation Analysis.} \sgta, and \mlsf have their corresponding selection and fetching policies. To understand how the learned selection and fetching policies $\Pi_{select}$ and $\Pi_{fetch}$ affect \sgta individually, we plug them one at a time. Table \ref{tbl:ablation} shows the results of this analysis for all the three datasets. If we employ $\Pi_{fetch}$ for adaptive fetching inside \sgta, we get a speedup of 1.86, 2.19, and 3.36 for \yago, \dbpedia, and \freebase respectively, without losing any accuracy. Therefore, when practitioner requires 100\% accuracy, this combination can be deployed. This also shows the usefulness of $\Pi_{fetch}$ alone. However, the overall performance of \sgta with $\Pi_{fetch}$ alone is much worse than \mlsf and shows the importance of $\Pi_{select}$. 

\begin{table}[h]
\centering
\small
\begin{tabular}{cc|c|c|c|}
\cline{3-5}
                                                     &           & \multicolumn{3}{c|}{Test dataset}           \\ \cline{3-5} 
                                                     &           & \yago        & \dbpedia     & \freebase     \\ \hline
\multicolumn{1}{|c|}
{\parbox[t]{2mm}{\multirow{4}{*}{\rotatebox[origin=c]{90}{Train}}}} & \yago     & \textbf{(3.71, 0.93)} & (3.88, 0.96) & (22, 0.92)    \\ \cline{2-5} 
\multicolumn{1}{|c|}{}                               & \dbpedia  & (4.30, 0.90) & \textbf{(5, 0.97)}    & (55.36, 0.89) \\ \cline{2-5} 
\multicolumn{1}{|c|}{}                               & \freebase & (4.31, 0.90) & (5.64, 0.90) & \textbf{(47.71, 0.96)} \\ \cline{2-5} 
\multicolumn{1}{|c|}{}                               & Combined  & (3.87, 0.92) & (4.27, 0.97) & (55.84, 0.96) \\ \hline
\end{tabular}
\caption{Transfer results (speedup, accuracy). }
\label{tbl:transferLearning}
\end{table}

\noindent {\bf Transfer Learning.} The learned policy can be seen as a function that maps search states to appropriate actions via features. Since the feature definitions are general, one could hypothesize that the learned knowledge is general and can be used to query different data graphs. To test this hypothesis, we learned policies on each of the three datasets, another policy using the combination of all the three datasets; and tested each policy on both individual datasets and the combined dataset. We make the following observations from Table \ref{tbl:transferLearning}. The learned policies generalize reasonably well to datasets that are not used for their training. We get the best accuracy when training and testing are done on the same dataset. The only exception is that the policy trained on the combined dataset gave better performance on \freebase. 

\begin{table}[h]
\centering
\small
\begin{tabular}{|c|c|}
\hline
Change \% & (Speedup, Accuracy) \\ \hline
0\%    & (4.12, 95.9\%)     \\ \hline
5\%    & (5.26, 93.2\%)     \\ \hline
10\%   & (7.03, 93.1\%)     \\ \hline
\end{tabular}
\caption{Performance with changes to the trained data graph.}
\label{tbl:dynamicGraph}
\end{table}

\noindent {\bf Changing Data Graph.} To investigate the stability of the learned policy with changes to the training data graph, we performed some experiments on \dbpedia graph. We do not have access to the temporal data graph. Therefore, we created multiple samples of original data graph with varying sizes by employing the well-studied {\em Forest Fire (FF)} approach~\cite{leskovec2006sampling}. We train on the smallest data sample (90\% of the original graph) and test on larger samples. This setup is based on the fact that real-world graphs are known to become large and dense over time~\cite{leskovec2007graph}. From Table~\ref{tbl:dynamicGraph}, we can see that by changing $10\%$ of the training data graph (i.e., 1.6M additional edges), \mlsf loses at most $3\%$ accuracy (similar to transfer learning results). Additionally, the speedup of \mlsf improves as the data graph evolves. As explained before, it is natural to expect more speedup with large and dense graphs: increased candidate matches for each sub-query can make \sgta slower and \mlsf can be more beneficial. Indeed, Table~\ref{tbl:highlevel} corroborates this hypothesis~\eg{ \freebase}.

In general, we need to update the policy whenever the distribution of queries and/or data graph changes significantly.

\section{Related Work}
\label{sec:related}

Our work is related to a sub-area of AI called speedup learning \cite{SL}. Specifically, it is an instance of inter-problem speedup learning. Reinforcement learning (RL), imitation learning (IL), and hybrid approaches combining RL and IL have been explored to learn search control knowledge from training problems in the context of diverse application domains. Some examples include job shop scheduling \cite{JS}, deterministic and stochastic planning \cite{DP,SP}, natural language processing \cite{nlp1,nlp2,nlp3,PS:EMNLP2014}, computer vision \cite{cv1,cv2}, hardware design optimization \cite{ICCAD15,TCAD16}, sensor data analysis \cite{AP:KDD2015}, and mixed-integer programming solvers \cite{mip1,mip2}. Our work explores this speedup learning problem for query planning in graph databases, a novel application domain. We formalized and solved this problem in a learning to plan framework. There is some work on applying learning in the context of query optimization \cite{hasan2014machine,ganapathi2009predicting,gupta2008pqr}, but we are not aware of any existing work that tightly integrates learning and search for graph query processing. 

Knoblock and Kambhampati has done seminal work on applying automating planning techniques for information integration on the web \cite{IIW-Tutorial}. Their work leverages the relationship between query planning in information integration and automated planning with sensing (i.e., information gathering) actions. Our work is different from theirs as we explore query planning in the context of traditional (graph) databases and rely heavily on advanced learning techniques.

\section{Summary and Future Work}
\label{sec-conclude}

We developed a general learning to plan (L2P) framework to improve the computational efficiency of a large-class of query reasoners that follow the Threshold Algorithm framework. 
We showed that our concrete instantiation \mlsf can achieve significant speedup over \sgta with negligible loss in accuracy across multiple knowledge graphs. Future work includes 
scaling up L2P framework to handle large number of batch/streaming queries by exploring active learning techniques; and deploying L2P instantiations for real-world relational and graph databases.

{\bf Acknowledgements.} {\footnotesize Jana Doppa would like to thank Tom Dietterich, Alan Fern, and Prasad Tadepalli for useful discussions. This work was supported in part by a Google Faculty Research award, and in part by the NSF grant \#1543656. }

\clearpage
\footnotesize
\bibliographystyle{aaai}
\bibliography{paper}

\begin{thebibliography}{}

\bibitem[\protect\citeauthoryear{Arai \bgroup et al\mbox.\egroup
  }{2007}]{Arai:2007:AMT:1325851.1325954}
Arai, B.; Das, G.; Gunopulos, D.; and Koudas, N.
\newblock 2007.
\newblock Anytime measures for top-k algorithms.
\newblock In {\em VLDB}.

\bibitem[\protect\citeauthoryear{Chen and Guestrin}{2016}]{XGBOOST}
Chen, T., and Guestrin, C.
\newblock 2016.
\newblock Xgboost: {A} scalable tree boosting system.
\newblock In {\em KDD}.

\bibitem[\protect\citeauthoryear{Chin~Jr \bgroup et al\mbox.\egroup
  }{2014}]{chin2014predicting}
Chin~Jr, G.; Choudhury, S.; Feo, J.; and Holder, L.
\newblock 2014.
\newblock Predicting and detecting emerging cyberattack patterns using
  streamworks.
\newblock In {\em Proceedings of the 9th Annual Cyber and Information Security
  Research Conference},  93--96.

\bibitem[\protect\citeauthoryear{Das \bgroup et al\mbox.\egroup
  }{2015}]{ICCAD15}
Das, S.; Doppa, J.~R.; Kim, D.; Pande, P.~P.; and Chakrabarty, K.
\newblock 2015.
\newblock Optimizing {3D} {NoC} design for energy efficiency: {A} machine
  learning approach.
\newblock In {\em ICCAD},  705--712.

\bibitem[\protect\citeauthoryear{Das \bgroup et al\mbox.\egroup }{2016a}]{db3}
Das, M.; Wu, Y.; Khot, T.; Kersting, K.; and Natarajan, S.
\newblock 2016a.
\newblock Scaling lifted probabilistic inference and learning via graph
  databases.
\newblock In {\em SDM}.

\bibitem[\protect\citeauthoryear{Das \bgroup et al\mbox.\egroup
  }{2016b}]{TCAD16}
Das, S.; Doppa, J.~R.; Pande, P.~P.; and Chakrabarty, K.
\newblock 2016b.
\newblock Design-space exploration and optimization of an energy-efficient and
  reliable {3D} small-world network-on-chip.
\newblock {\em TCAD}.

\bibitem[\protect\citeauthoryear{Ding \bgroup et al\mbox.\egroup
  }{2014}]{ding2014top}
Ding, X.; Jia, J.; Li, J.; Liu, J.; and Jin, H.
\newblock 2014.
\newblock Top-k similarity matching in large graphs with attributes.
\newblock In {\em DASFAA}.

\bibitem[\protect\citeauthoryear{Doppa \bgroup et al\mbox.\egroup
  }{2014}]{MLS:AAAI14}
Doppa, J.~R.; Yu, J.; Ma, C.; Fern, A.; and Tadepalli, P.
\newblock 2014.
\newblock {HC}-{S}earch for {M}ulti-{L}abel {P}rediction: {A}n {E}mpirical
  {S}tudy.
\newblock In {\em AAAI}.

\bibitem[\protect\citeauthoryear{Doppa, Fern, and Tadepalli}{2014a}]{JAIR14}
Doppa, J.~R.; Fern, A.; and Tadepalli, P.
\newblock 2014a.
\newblock {HC}-{S}earch: A learning framework for search-based structured
  prediction.
\newblock {\em JAIR} 50:369--407.

\bibitem[\protect\citeauthoryear{Doppa, Fern, and Tadepalli}{2014b}]{JMLR14}
Doppa, J.~R.; Fern, A.; and Tadepalli, P.
\newblock 2014b.
\newblock Structured prediction via output space search.
\newblock {\em JMLR} 15:1317--1350.

\bibitem[\protect\citeauthoryear{Fagin, Lotem, and
  Naor}{2003}]{fagin2003optimal}
Fagin, R.; Lotem, A.; and Naor, M.
\newblock 2003.
\newblock Optimal aggregation algorithms for middleware.
\newblock {\em Journal of Computer and System Sciences} 66(4):614--656.

\bibitem[\protect\citeauthoryear{Fern, Yoon, and Givan}{2006}]{API}
Fern, A.; Yoon, S.~W.; and Givan, R.
\newblock 2006.
\newblock Approximate policy iteration with a policy language bias: Solving
  relational markov decision processes.
\newblock {\em JAIR} 25:75--118.

\bibitem[\protect\citeauthoryear{Fern}{2010}]{SL}
Fern, A.
\newblock 2010.
\newblock Speedup learning.
\newblock In {\em Encyclopedia of Machine Learning}.

\bibitem[\protect\citeauthoryear{Fern}{2016}]{PC}
Fern, A.
\newblock 2016.
\newblock Personal Communication.

\bibitem[\protect\citeauthoryear{Ganapathi \bgroup et al\mbox.\egroup
  }{2009}]{ganapathi2009predicting}
Ganapathi, A.; Kuno, H.; Dayal, U.; Wiener, J.~L.; Fox, A.; Jordan, M.; and
  Patterson, D.
\newblock 2009.
\newblock Predicting multiple metrics for queries: Better decisions enabled by
  machine learning.
\newblock In {\em ICDE}.

\bibitem[\protect\citeauthoryear{Gupta, Mehta, and Dayal}{2008}]{gupta2008pqr}
Gupta, C.; Mehta, A.; and Dayal, U.
\newblock 2008.
\newblock Pqr: Predicting query execution times for autonomous workload
  management.
\newblock In {\em ICAC}.

\bibitem[\protect\citeauthoryear{Hasan and Gandon}{2014}]{hasan2014machine}
Hasan, R., and Gandon, F.
\newblock 2014.
\newblock A machine learning approach to sparql query performance prediction.
\newblock In {\em IEEE/WIC/ACM}.

\bibitem[\protect\citeauthoryear{He, {Daum{\'{e}} III}, and
  Eisner}{2012}]{nlp2}
He, H.; {Daum{\'{e}} III}, H.; and Eisner, J.
\newblock 2012.
\newblock Imitation learning by coaching.
\newblock In {\em NIPS}.

\bibitem[\protect\citeauthoryear{He, {Daum{\'{e}} III}, and
  Eisner}{2013}]{nlp3}
He, H.; {Daum{\'{e}} III}, H.; and Eisner, J.
\newblock 2013.
\newblock Dynamic feature selection for dependency parsing.
\newblock In {\em EMNLP}.

\bibitem[\protect\citeauthoryear{He, {Daum{\'{e}} III}, and
  Eisner}{2014}]{mip1}
He, H.; {Daum{\'{e}} III}, H.; and Eisner, J.
\newblock 2014.
\newblock Learning to search in branch and bound algorithms.
\newblock In {\em NIPS}.

\bibitem[\protect\citeauthoryear{Jiang \bgroup et al\mbox.\egroup
  }{2012}]{nlp1}
Jiang, J.; Teichert, A.~R.; {Daum{\'{e}} III}, H.; and Eisner, J.
\newblock 2012.
\newblock Learned prioritization for trading off accuracy and speed.
\newblock In {\em NIPS}.

\bibitem[\protect\citeauthoryear{Khalil \bgroup et al\mbox.\egroup
  }{2016}]{mip2}
Khalil, E.~B.; Bodic, P.~L.; Song, L.; Nemhauser, G.~L.; and Dilkina, B.~N.
\newblock 2016.
\newblock Learning to branch in mixed integer programming.
\newblock In {\em AAAI}.

\bibitem[\protect\citeauthoryear{Kim \bgroup et al\mbox.\egroup
  }{2015}]{kim2015taming}
Kim, J.; Shin, H.; Han, W.-S.; Hong, S.; and Chafi, H.
\newblock 2015.
\newblock Taming subgraph isomorphism for rdf query processing.
\newblock {\em VLDB}  1238--1249.

\bibitem[\protect\citeauthoryear{Knoblock and Kambhampati}{2007}]{IIW-Tutorial}
Knoblock, C., and Kambhampati, S.
\newblock 2007.
\newblock Tutorial on information integration on the web.
\newblock In {\em AAAI}.

\bibitem[\protect\citeauthoryear{Lam \bgroup et al\mbox.\egroup
  }{2015}]{CVPR15}
Lam, M.; Doppa, J.~R.; Todorovic, S.; and Dietterich, T.
\newblock 2015.
\newblock {HC}-{S}earch for structured prediction in computer vision.
\newblock In {\em CVPR}.

\bibitem[\protect\citeauthoryear{Leskovec and
  Faloutsos}{2006}]{leskovec2006sampling}
Leskovec, J., and Faloutsos, C.
\newblock 2006.
\newblock Sampling from large graphs.
\newblock In {\em SIGKDD}.

\bibitem[\protect\citeauthoryear{Leskovec, Kleinberg, and
  Faloutsos}{2007}]{leskovec2007graph}
Leskovec, J.; Kleinberg, J.; and Faloutsos, C.
\newblock 2007.
\newblock Graph evolution: Densification and shrinking diameters.
\newblock {\em TKDD} ~2.

\bibitem[\protect\citeauthoryear{Lu \bgroup et al\mbox.\egroup
  }{2013}]{LuLWLW13}
Lu, J.; Lin, C.; Wang, W.; Li, C.; and Wang, H.
\newblock 2013.
\newblock String similarity measures and joins with synonyms.
\newblock In {\em SIGMOD}.

\bibitem[\protect\citeauthoryear{Ma \bgroup et al\mbox.\egroup
  }{2014}]{PS:EMNLP2014}
Ma, C.; Doppa, J.~R.; Orr, J.~W.; Mannem, P.; Fern, X.~Z.; Dietterich, T.~G.;
  and Tadepalli, P.
\newblock 2014.
\newblock Prune-and-score: Learning for greedy coreference resolution.
\newblock In {\em EMNLP}.

\bibitem[\protect\citeauthoryear{Martinez-Cantin}{2014}]{JMLR:v15:martinezcantin14a}
Martinez-Cantin, R.
\newblock 2014.
\newblock Bayesopt: A bayesian optimization library for nonlinear optimization,
  experimental design and bandits.
\newblock {\em Journal of Machine Learning Research} 15:3915--3919.

\bibitem[\protect\citeauthoryear{Minor, Doppa, and Cook}{2015}]{AP:KDD2015}
Minor, B.; Doppa, J.~R.; and Cook, D.~J.
\newblock 2015.
\newblock Data-driven activity prediction: Algorithms, evaluation methodology,
  and applications.
\newblock In {\em KDD}.

\bibitem[\protect\citeauthoryear{Morsey \bgroup et al\mbox.\egroup
  }{2011}]{MOR+11}
Morsey, M.; Lehmann, J.; Auer, S.; and {Ngonga Ngomo}, A.-C.
\newblock 2011.
\newblock {DB}pedia {SPARQL} {B}enchmark -- {P}erformance {A}ssessment with
  {R}eal {Q}ueries on {R}eal {D}ata.
\newblock In {\em ISWC}.

\bibitem[\protect\citeauthoryear{Navarro}{2001}]{navarro2001guided}
Navarro, G.
\newblock 2001.
\newblock A guided tour to approximate string matching.
\newblock {\em CSUR} 33(1):31--88.

\bibitem[\protect\citeauthoryear{Niu \bgroup et al\mbox.\egroup }{2011}]{db1}
Niu, F.; R{\'{e}}, C.; Doan, A.; and Shavlik, J.~W.
\newblock 2011.
\newblock Tuffy: Scaling up statistical inference in markov logic networks
  using an {RDBMS}.
\newblock {\em {PVLDB}} 4(6):373--384.

\bibitem[\protect\citeauthoryear{Pinto and Fern}{2014}]{SP}
Pinto, J., and Fern, A.
\newblock 2014.
\newblock Learning partial policies to speedup {MDP} tree search.
\newblock In {\em UAI}.

\bibitem[\protect\citeauthoryear{Prud’Hommeaux, Seaborne, and
  others}{2008}]{prud2006sparql}
Prud’Hommeaux, E.; Seaborne, A.; et~al.
\newblock 2008.
\newblock Sparql query language for rdf.
\newblock {\em W3C recommendation} 15.

\bibitem[\protect\citeauthoryear{Ross, Gordon, and Bagnell}{2011}]{DAGGER}
Ross, S.; Gordon, G.~J.; and Bagnell, D.
\newblock 2011.
\newblock A reduction of imitation learning and structured prediction to
  no-regret online learning.
\newblock In {\em AISTATS}.

\bibitem[\protect\citeauthoryear{Roy, Eliassi-Rad, and
  Papadimitriou}{2015}]{6877688}
Roy, S.~B.; Eliassi-Rad, T.; and Papadimitriou, S.
\newblock 2015.
\newblock Fast best-effort search on graphs with multiple attributes.
\newblock {\em TKDE} 27(3):755--768.

\bibitem[\protect\citeauthoryear{Sarkhel \bgroup et al\mbox.\egroup
  }{2016}]{db2}
Sarkhel, S.; Venugopal, D.; Pham, T.~A.; Singla, P.; and Gogate, V.
\newblock 2016.
\newblock Scalable training of markov logic networks using approximate
  counting.
\newblock In {\em AAAI}.

\bibitem[\protect\citeauthoryear{Shahriari \bgroup et al\mbox.\egroup
  }{2016}]{BO}
Shahriari, B.; Swersky, K.; Wang, Z.; Adams, R.~P.; and de~Freitas, N.
\newblock 2016.
\newblock Taking the human out of the loop: {A} review of bayesian
  optimization.
\newblock {\em Proceedings of the {IEEE}} 104(1):148--175.

\bibitem[\protect\citeauthoryear{Weiss and Taskar}{2013}]{cv2}
Weiss, D.~J., and Taskar, B.
\newblock 2013.
\newblock Learning adaptive value of information for structured prediction.
\newblock In {\em NIPS}.

\bibitem[\protect\citeauthoryear{Weiss, Sapp, and Taskar}{2013}]{cv1}
Weiss, D.~J.; Sapp, B.; and Taskar, B.
\newblock 2013.
\newblock Dynamic structured model selection.
\newblock In {\em ICCV}.

\bibitem[\protect\citeauthoryear{Xu, Fern, and Yoon}{2009}]{DP}
Xu, Y.; Fern, A.; and Yoon, S.~W.
\newblock 2009.
\newblock Learning linear ranking functions for beam search with application to
  planning.
\newblock {\em JMLR} 10:1571--1610.

\bibitem[\protect\citeauthoryear{Yang \bgroup et al\mbox.\egroup
  }{2016}]{yangfast}
Yang, S.; Han, F.; Wu, Y.; and Yan, X.
\newblock 2016.
\newblock Fast top-k search in knowledge graphs.
\newblock In {\em ICDE}.

\bibitem[\protect\citeauthoryear{Zeng \bgroup et al\mbox.\egroup
  }{2012}]{zeng2012top}
Zeng, X.; Cheng, J.; Yu, J.; and Feng, S.
\newblock 2012.
\newblock Top-k graph pattern matching: A twig query approach.
\newblock {\em WAIM}.

\bibitem[\protect\citeauthoryear{Zhang and Dietterich}{1995}]{JS}
Zhang, W., and Dietterich, T.~G.
\newblock 1995.
\newblock A reinforcement learning approach to job-shop scheduling.
\newblock In {\em IJCAI}.

\bibitem[\protect\citeauthoryear{Zhang, Kumar, and R{\'{e}}}{2016}]{db4}
Zhang, C.; Kumar, A.; and R{\'{e}}, C.
\newblock 2016.
\newblock Materialization optimizations for feature selection workloads.
\newblock {\em ACM TODS} 41(1):2.

\bibitem[\protect\citeauthoryear{Zou \bgroup et al\mbox.\egroup
  }{2014}]{ZouHWYHZ14}
Zou, L.; Huang, R.; Wang, H.; Yu, J.~X.; He, W.; and Zhao, D.
\newblock 2014.
\newblock Natural language question answering over {RDF:} a graph data driven
  approach.
\newblock In {\em SIGMOD}.

\end{thebibliography}

\end{document}